\documentclass[12pt]{iopart}

\usepackage{graphicx}
\usepackage{color}
\usepackage{braket}
\usepackage{cite}
\usepackage{dsfont}
\usepackage{amssymb}
\usepackage{mathabx}

\usepackage{hyperref}
\hypersetup{
    colorlinks,
    citecolor=blue,
    filecolor=red,
    linkcolor=magenta,
    urlcolor=red
}

\begin{document}

\title{Laser-free trapped ion entangling gates with AESE: \\ Adiabatic Elimination of Spin-motion Entanglement}

\author{R. Tyler Sutherland$^{1,2}$, Michael Foss-Feig $^{1}$}

\address{$^1$ Quantinuum, 303 S Technology Ct, Broomfield, CO 80021, USA}
\address{$^2$ Department of Electrical and Computer Engineering, University of Texas at San Antonio, San Antonio, Texas 78249, USA}

\ead{robert.sutherland@quantinuum.com}

\begin{abstract}
We discuss a laser-free, two-qubit geometric phase gate technique for generating high-fidelity entanglement between two trapped ions. The scheme works by ramping the spin-dependent force on and off slowly relative to the gate detunings, which adiabatically eliminates the spin-motion entanglement (AESE). We show how gates performed with AESE can eliminate spin-motion entanglement with multiple modes simultaneously, without having to specifically tune the control field detunings. This is because the spin-motion entanglement is suppressed by operating the control fields in a certain parametric limit,  rather than by engineering an optimized control sequence. We also discuss physical implementations that use either electronic or ferromagnetic magnetic field gradients. In the latter, we show how to ``AESE" the system by smoothly turning on the \textit{effective} spin-dependent force by shelving from a magnetic field insensitive state to a magnetic field sensitive state slowly relative to the gate mode frequencies. We show how to do this with a Rabi or adiabatic rapid passage transition. Finally, we show how gating with AESE significantly decreases the gate's sensitivity to common sources of motional decoherence, making it easier to perform high-fidelity gates at Doppler temperatures.
\end{abstract}

\maketitle
\section{Introduction}

Trapped ions are amongst the most promising candidate platforms for scalable quantum computing, affording high fidelity gates, high connectivity, low cross-talk, and long coherence times \cite{cirac_1995, monroe_1995, wineland_1998, haffner_2008,blatt_2008,ladd_2010, harty_2014,ballance_2016,gaebler_2016,bruzewicz_2019,srinivas_2021,clark_2021,moses_2023}. While every proposal for scaling trapped ion quantum computers to a size capable of useful computations comes with its own technical challenges, maintaining sufficiently high two-qubit gate fidelities is always among the most significant. First of all, no one has generated two-qubit gates with infidelities well-below the $\sim 10^{-3}$ threshold required by many approaches to quantum error correction \cite{campbell_2017}. While two qubit error rates at the level of $\sim 10^{-4}$ seem feasible, it will be another challenge entirely to maintain such unprecedented fidelities while scaling to systems with many qubits. It is therefore desirable to explore new gating schemes, evaluating them based on their potential for high-fidelity operations and scalability. \\

\noindent Most contemporary trapped ion two-qubit gate schemes are, fundamentally, geometric phase gates \cite{molmer_1999,molmer_2000,leibfried_2003}. Their ubiquity stems from their relaxed mode temperature requirements; the simplest model of a geometric phase gate incurs no direct errors from operating at non-zero temperatures. There are, however, several sources of infidelity stemming from deviations from this idealized model which worsen with temperature, leading the highest-fidelity gate demonstration thus far to leverage ground state cooling \cite{gaebler_2016,ballance_2016,srinivas_2021,clark_2021}. Something that briefly appeared to have the potential to further decrease temperature requirements was the concept of the ``push gate" \cite{cirac_2000,calarco_2001,savsura_2003}; for push gates, it was proposed to operate in the far-detuned limit described originally by M\o lmer and S\o rensen \cite{molmer_1999}, further eliminating spin-motion entanglement by adiabatically ramping on/off the spin-dependent force. In this limit, the final propagator for the push gate only depends on the mean displacement of the atomic wavepacket, ostensibly negating the scheme's temperature dependence. Because that work focused primarily on lasers, however, its potential use in high-fidelity operations was limited by photon scattering, and the temperature insensitivity noted in Ref.~\cite{cirac_2000} was limited by higher-order Lamb-Dicke effects \cite{savsura_2003}. Crucially, both photon scattering and higher-order Lamb-Dicke effects are negligible in systems that rely on spin-dependent forces generated by magnetic field gradients instead of lasers, \cite{ospelkaus_2008,ospelkaus_2011,harty_2016,weidt_2016,webb_2018,srinivas_2018,sutherland_2019,srinivas_2021}, suggesting the scheme should be reconsidered in this `laser-free' context. \\

\noindent One of the best arguments \emph{against} working with gradients that are far detuned from their respective gating modes is that---all else being equal---doing so decreases gate speed; this decrease seems especially problematic for laser-free gates, as they are already slow relative to their laser counterparts. There are some problems with this argument, however. The first is the assumption that far-detuned gates must be performed in the same parameter regime as near-detuned gates, which is not necessarily valid. As we will show, far-detuned gates can be made extremely insensitive to motional errors. This insensitivity could enable operation in regimes where we have significantly increased the Lamb-Dicke factor of certain modes\textemdash for example, by operating near the zig-zag instability of a two-ion crystal\textemdash which could result in faster laser-free gates than the state-of-the art (see Sec.~\ref{sec:rf}). Secondly, in QCCD architectures, gate times currently occupy only a small fraction of the overall time budget, which is typically dominated by transport and ground-state cooling \cite{pino_2021}. As we discuss below, far-detuned gates could decrease the time required for both of these tasks, potentially decreasing algorithm times much more than what could be done via speeding up two-qubit gates by themselves. The remaining concern with (potentially) increasing gate times via operating in the far-detuned regime is a correspondent increase in error; below, we argue that far-detuned gates can be operated such that the opposite is true, at least for motional errors\textemdash the largest source of infidelity in laser-free gates \cite{srinivas_2021}. \\

\noindent Traditionally, geometric phase gates operate with an effective spin-dependent force tuned near-resonant to one (or more) of the ion chain's motional modes. This makes the gate faster \cite{molmer_1999,molmer_2000}, but also necessitates a control sequence designed so there is no spin-motion entanglement at the end of the gate. Any control sequence must then be designed with the \textit{specific} set of system parameters (mode frequencies, mode structure, etc..) taken into account, meaning, if these parameters fluctuate, there will be residual spin-motion entanglement. In this paper, we discuss an alternative method that suppresses spin-motion entanglement when the gate is operated in a \textit{limit} where the spin-dependent force is smoothly ramped on/off slowly relative to the gate mode detunings. This procedure adiabatically eliminates the spin-motion entanglement AESE (pronounced `ease'), achieving significant suppression of residual spin-motion entanglement using a technique that is agnostic to specific values of system parameters. The fact that AESE depends only on a parametric limit, rather than a specific set of parameters, makes the probability of its success high. Further, the relative simplicity of AESE\textemdash where the only free parameter is the gradient ramp time\textemdash is worth noting when contrasted with the sophisticated dynamical decoupling sequences used to suppress spin-motion entanglement previously \cite{hayes_2012,arrazola_2018,srinivas_2021,barthel_2022}. Finally, the two major sources of error reported in Ref.~\cite{srinivas_2021}, currently the highest fidelity laser-free two-qubit gate demonstration, were heating and mode frequency fluctuations. As shown in Ref.~\cite{sutherland_2022_1}, only the latter of these errors increases with mode temperature, indicating that gates performed with AESE could maintain high fidelities at non-zero temperatures.\\

\noindent In order to perform a geometric phase gate with AESE, it is necessary to turn on and off a spin-dependent force smoothly and slowly. This is straightforward if a gate is driven by electronic gradients, see Sec.~\ref{sec:electronic}, as we can simply ramp the gradient generating current. If the gradient cannot be controlled because it is made with a permanent ferromagnet, however, the only option is to smoothly/slowly ramp the ion's response to the gradient. One way of doing this is by shelving from a magnetic field insensitive `clock' state, into a magnetic field sensitive `Zeeman state', enabling a ramp of the \textit{effective} spin-dependent force over a predetermined time. In Sec.~\ref{sec:ferromagnets}, we compare the performance when doing this via a Rabi or adiabatic rapid passage (ARP) transition.

\section{Theory}

Geometric phase gates take many forms and can be used to generate spin \cite{katz_2022} as well as phonon \cite{sutherland_2021_cvqc} interactions beyond $2^{\mathrm{nd}}$ order. In this work, however, we consider a subset of geometric phase gates typically used for high-fidelity two-qubit gates, which we represent with the Hamiltonian:
\begin{eqnarray}\label{eq:first_geo}
\hat{H}_{\mathrm{g}} = \hbar\gamma(t)\sum_{j}\cos(\omega_{g}t)\Omega_{j}\hat{S}_{z,j}\Big(\hat{a}_{j}^{\dagger}e^{i\omega_{j}t} + \hat{a}_{j}e^{-i\omega_{j}t} \Big).
\end{eqnarray}
Here, $\hat{S}_{z,j}\equiv \sum_{k}b^{j}_{k}\hat{\sigma}_{z,k}$ is the collective Pauli-$z$ operator associated with gating mode $j$ of a same- or mixed-species crystal; for example, $\hat{S}_{z,\mathrm{c}}\equiv \hat{\sigma}_{z,1}+\hat{\sigma}_{z,2}$ for the center-of-mass mode and $\hat{S}_{z,\mathrm{s}}\equiv \hat{\sigma}_{z,1}-\hat{\sigma}_{z,2}$ for the stretch (out-of-phase) mode in a same-species two ion crystal. We have chosen a gate comprising Pauli-z operators, but the conclusions we draw about AESE are trivially generalized to gates with Pauli operators on the xy-plane \cite{ospelkaus_2011,harty_2016,zarantonello_2019}. We consider cases when the gradient frequency $\omega_{g}$ is DC ($\omega_{g}=0$) or RF ($\omega_{g}\sim \mathrm{MHz}$). The value of $\omega_{j}$ represents the frequency of the crystal's $j^{\mathrm{th}}$ motional mode. The function $\gamma(t)$ represents the time dependence of the gradient envelope, assumed to smoothly increase(decrease) from $0(1)$ to $1(0)$ over a time $\tau$ at the beginning(end) of the operation. The sideband of mode $j$ has a Rabi frequency of $\Omega_{j}$, and is given by:
\begin{eqnarray}
    \Omega_{j} \equiv \frac{\tilde{r}_{j}(\nabla B_{\mathrm{q}}\cdot \hat{r}_{j})}{2}\frac{\partial\omega_{0}}{\partial B_{q}},
\end{eqnarray}
where $\hat{r}_{j}$ is a unit vector along the principal axis of mode $j$, $\tilde{r}_{j}\equiv \sqrt{\hbar/2\omega_{j}m_{j}}$ is the mode's spatial extent at its ground state energy, and $m_{j}$ its effective mass. We represent the qubit frequency as $\omega_{0}$, and the magnetic field's projection along the quantization axis as $B_{\mathrm{q}}$. We can generate this type of state-dependent force with the near-field \cite{leibfried_2007} or far-field \cite{mintert_2001,weidt_2016,webb_2018} of a permanent magnet, or with the near-field of an integrated circuit \cite{wineland_1998,ospelkaus_2008,ospelkaus_2011,harty_2016,srinivas_2018, zarantonello_2019, srinivas_2021}. Finally, in Eq.~(\ref{eq:first_geo}), the collective spin degrees of freedom $\hat{S}_{z,j}$ are not directly coupled, mutually commute, and each couples to an independent motional mode. As a result, the time propagator factors into contributions from each $\hat{S}_{z,j}$. Therefore, each collective spin's contribution to the total entanglement angle $\phi$ is additive; we will often consider one mode at a time below, dropping the subscript $j$. \\

\noindent We analyze the gate dynamics using the Magnus \cite{magnus_1954} expansion up to $2^{\mathrm{nd}}$:
\begin{eqnarray}\label{eq:magnus}
    \hat{U}_{\mathrm{g}} = \exp\Big(-\frac{i}{\hbar}\int^{t_{g}}_{0}dt^{\prime}\hat{H}_{\mathrm{g}}(t^{\prime}) - \frac{1}{\hbar^{2}}\int^{t_{g}}_{0}\int^{t^{\prime}}_{0}dt^{\prime}dt^{\prime\prime}\Big[\hat{H}_{\mathrm{g}}(t^{\prime}),\hat{H}_{\mathrm{g}}(t^{\prime\prime}) \Big] \Big),
\end{eqnarray}
which gives the \textit{exact} time propagator for Eq.~(\ref{eq:first_geo}) for a gate duration $t_{g}$. Of the two terms in our expression, the $1^{\mathrm{st}}$ order terms represent spin-motion coupling, while the $2^{\mathrm{nd}}$ order terms represent spin-spin coupling. Any high-fidelity gate scheme based on Eq.~(\ref{eq:first_geo}) has to ensure the $1^{\mathrm{st}}$ order terms become sufficiently small, and the $2^{\mathrm{nd}}$ order terms accumulate phases that add to $\phi$. We now focus on the former requirement.

\subsection{Adiabatic Elimination of the Spin-motion Entanglement}\label{sec:u1}

We can factor $\hat{U}_{g}$ into two unitaries that represent the $1^{\mathrm{st}}$ and $2^{\mathrm{nd}}$ order terms of the time propagator because the two terms commute. We therefore consider the $1^{\mathrm{st}}$ order terms in the Magnus expansion separately:
\begin{eqnarray}\label{eq:spin_mot_int}
    \hat{U}_{1} = \exp\Big(-i\Omega_{\mathrm{g}}^{\prime}\hat{S}_{z}\int^{t_{g}}_{0}dt^{\prime}\gamma(t^{\prime}) \Big[\hat{a}^{\dagger}e^{i\delta t^{\prime}} + \hat{a}e^{-i\delta t^{\prime}}\Big]\Big),
\end{eqnarray}
where $\delta\equiv \omega\pm\omega_{\mathrm{g}}$ represents the field detuning from a `gating' motional mode and $\Omega_{\mathrm{g}}^{\prime}\equiv \Omega_{\mathrm{g}}/2$ or $\Omega_{\mathrm{g}}^{\prime}\equiv\Omega_{\mathrm{g}}$, depending on if the gradient is DC ($\omega_{\mathrm{g}}=0$) or RF ($\omega_{\mathrm{g}}\neq 0$), respectively. We can simplify this equation via integration by parts:
\begin{eqnarray}
\label{eq:mag_parts}
\hat{U}_{1} &=  \exp\!\Big(\frac{\Omega_{\mathrm{g}}^{\prime}}{\delta}\hat{S}_{z}\Big[\int^{\tau}_{0}\!\!\!dt^{\prime}\dot{\gamma}(t^{\prime})(\hat{a}^{\dagger}e^{i\delta t^{\prime}} - \hat{a}e^{-i\delta t^{\prime}})\!+ \!\!\int^{t_{g}}_{t_{g}-\tau}\!\!\!\!\!\!\!\!dt^{\prime}\dot{\gamma}(t^{\prime})(\hat{a}^{\dagger}e^{i\delta t^{\prime}} - \hat{a}e^{-i\delta t^{\prime}})\Big] \Big) \nonumber \\
&=\exp\!\Big(\frac{i\Omega_{\mathrm{g}}^{\prime}}{\delta^{2}}\hat{S}_{z}\Big[\int^{\tau}_{0}\!\!\!dt^{\prime}\ddot{\gamma}(t^{\prime})(\hat{a}^{\dagger}e^{i\delta t^{\prime}} + \hat{a}e^{-i\delta t^{\prime}}) \!+ \!\!\int^{t_{g}}_{t_{g}-\tau}\!\!\!\!\!\!\!\!dt^{\prime}\ddot{\gamma}(t^{\prime})(\hat{a}^{\dagger}e^{i\delta t^{\prime}} + \hat{a}e^{-i\delta t^{\prime}})\Big] \Big), \nonumber \\
\end{eqnarray}
where we have assumed that $\gamma(0)=\dot{\gamma}(0)=\dot{\gamma}(\tau)=\dot{\gamma}(t_{g}-\tau)=\gamma(t_{g})=\dot{\gamma}(t_{g})=0$, and $\dot{\gamma}(t)=0$ when $t\in[\tau,t_{g}-\tau]$. Under these assumptions, the off-resonant couplings in Eq.~(\ref{eq:mag_parts}) will be suppressed by pulse shaping $\gamma(t)$, similar to the way this is done to suppress off-resonant Rabi flopping on the carrier transition \cite{roos_2008,sutherland_2019}. In general, the magnitude of $\Omega_{\mathrm{g}}^{\prime}\delta^{-2}\int^{t}_{t_{0}}\ddot\gamma(t)e^{\pm i\delta t}~\propto~\Omega_{\mathrm{g}}^{\prime}/(\delta^{3}\tau^{2})$, meaning the infidelity from residual spin-motion entanglement (per mode) will scale like $\Omega_{\mathrm{g}}^{\prime 2}/(\delta^{6}\tau^{4})$. Choosing these properties of $\gamma(t)$ adiabatically eliminates the spin-motion entanglement (AESEs), suppressing its effect by roughly $(\delta\tau)^4$ compared to when $\tau\rightarrow 0$ and $\gamma(t)\rightarrow 1$. The ions must complete several phase space loops over $t_{g}$, but this additional suppression marks a crucial distinction between using AESE and simply operating in the `far-detuned' limit; the latter would require a significantly larger $\delta$ to achieve the same degree of spin-motion disentanglement, resulting in a much slower gate in comparison. We can factor $\hat{U}_{1}$ from the (assumed to be ideal) gate $\hat{U}_{2}=\hat{U}_{g}$, and evaluate the infidelity of a pair of qubits in the state $\ket{\psi(0)}$, initialized to the $n^{\mathrm{th}}$ Fock state:
\begin{eqnarray}\label{eq:infidelity}
    \mathcal{I}_{n} &=& 1-\sum_{n^{\prime}}|\bra{\psi(0)}\bra{n^{\prime}}\hat{U}_{\mathrm{1}}\ket{\psi(0)}\ket{n}|^{2}.
\end{eqnarray}
\\

\noindent For the specific examples in this work, we set $\gamma(t)\equiv \sin^{2}(\pi t/\tau)$. Plugging this expression into Eq.~(\ref{eq:mag_parts}), we get a spin-dependent displacement operator:
\begin{eqnarray}
    \hat{U}_{1} &=& \exp\Big(\hat{S}_{z}\frac{-i\pi^{2}\Omega_{\mathrm{g}}^{\prime}}{\delta(\pi^{2}-\tau^{2}\delta^{2})}\sin\Big[\delta[t_{g}-\tau]\Big]\Big[\hat{a}^{\dagger}e^{i\delta t_{g}/2} + \hat{a}e^{-i\delta t_{g}/2}\Big] \Big),
\end{eqnarray}
Inserting this into Eq.~(\ref{eq:infidelity}), assuming the limit $\tau\delta \gg 1$ has been reached, the time-averaged integrals are:
\begin{eqnarray}
    \mathcal{I}_{n} &\simeq & \frac{\pi^{4}\Omega_{\mathrm{g}}^{\prime 2}}{2\tau^{4}\delta^{6}}\Big(2n+1\Big)\lambda_{\hat{S}_{z}}^{2} \nonumber \\ 
    &= & \frac{2\pi^{4}\Omega_{\mathrm{g}}^{\prime 2}}{3\tau^{4}\delta^{6}}\Big(2n+1\Big)
\end{eqnarray}
where $\lambda_{\hat{S}_{z}}^{2}$ is the variance of $\hat{S}_{z}$ at $t=0$, which we take to be $8/5$ in the second line upon averaging over $\mathrm{SU}(4)$. \\

\noindent Traditionally, a geometric phase gate ensures spin-motion disentanglement by setting $t_{g}$ to be an integer multiple of $2\pi/\delta$. With AESE, however, we reach the same end by operating in a parametric limit, independent of set relationship between $\delta$ and $t_{g}$. One immediate result of this is that the mode frequencies can fluctuate significantly without inducing residual spin-motion entanglement, negating the need to increase experimental complexity with dynamical decoupling sequences \cite{hayes_2012, haddadfarshi_2016,webb_2018,zarantonello_2019,srinivas_2021}, and reducing temperature requirements  (see Sec.~\ref{sec:motional_decoherence}). Further, the gradient can now generate spin-motion coupling with many modes at once without the need for numerically optimized pulse sequences that ensure each qubit becomes disentangled with each mode at $t_{g}$ \cite{arrazola_2018,schafer_2018,barthel_2022}.

\subsection{Entanglement via $2^{\mathrm{nd}}$ order processes $\hat{U}_{2}$}

Similar to how shaping $\gamma(t)$ suppresses off-resonant $1^{\mathrm{st}}$ order transitions in the $\delta\tau\gg 1$ limit, it also suppresses off-resonant $2^{\mathrm{nd}}$ order transitions. To show this, we note Eq.~(\ref{eq:first_geo}) comprises a linear sum of terms, each with a time dependence $\propto~\gamma(t)e^{\pm i\delta t}$ for some angular frequency $\delta \equiv \omega\pm \omega_{g}$. We assume $|\delta|\tau\gg 1$ and that $|\delta^{\prime}-\delta^{\prime\prime}|\tau\gg 1$, where  the primes indicate different sidebands. Because of its $\propto [\hat{H}_{g}(t^{\prime}),\hat{H}_{g}(t^{\prime\prime})]$ dependence, the exponent of $\hat{U}_{2}$ is a linear sum of terms that are proportional to the time integral:
\begin{eqnarray}\label{eq:beta_orig}
    \beta(t_{g}) &=& \int^{t_{g}}_{0}\int^{t^{\prime}}_{0}dt^{\prime}dt^{\prime\prime}\gamma(t^{\prime})\gamma(t^{\prime\prime})\sin\Big(\delta^{\prime}t^{\prime}+\delta^{\prime\prime}t^{\prime\prime}\Big) \nonumber \\
    &=& -\int^{t_{g}}_{0}dt^{\prime}\gamma(t^{\prime})\Big\{\frac{\gamma(t^{\prime})}{\delta^{\prime\prime}}\cos([\delta^{\prime}+\delta^{\prime\prime}]t^{\prime})-\frac{1}{\delta^{\prime\prime}}\int^{t^{\prime}}_{0}dt^{\prime\prime}\dot{\gamma}(t^{\prime\prime})\cos(\delta^{\prime}t^{\prime}+\delta^{\prime\prime}t^{\prime\prime})\Big\},\nonumber \\
    &=& -\int^{t_{g}}_{0}dt^{\prime}\gamma(t^{\prime})\Big\{\frac{\gamma(t^{\prime})}{\delta^{\prime\prime}}\cos([\delta^{\prime}+\delta^{\prime\prime}]t^{\prime}) - \frac{\dot{\gamma}(t^{\prime})}{\delta^{\prime\prime 2}}\sin\Big([\delta^{\prime}+\delta^{\prime\prime}]t^{\prime} \Big) \nonumber \\ 
    &&~~~~~~~~~~~~~~~~~~~~+ \frac{1}{\delta^{\prime\prime 2}}\int^{t^{\prime}}_{0}dt^{\prime\prime}\ddot\gamma(t^{\prime\prime})\sin\Big(\delta^{\prime} t^{\prime}+\delta^{\prime\prime}t^{\prime\prime} \Big)\Big\}
\end{eqnarray}
where we have integrated by parts and used the fact that $\gamma(0)=\dot{\gamma}(0)=0$. When $|\delta^{\prime}-\delta^{\prime\prime}|\tau \gg 1$, every term in this equation scales $\propto~ \delta^{\prime\prime 2}\tau$, their effect on the gate fidelity (again) scaling $\propto ~\Omega_{\mathrm{g}}^{\prime 2}/(\delta^{6}\tau^{4})$. For this reason, we assume all terms such that $\delta^{\prime}\neq \delta^{\prime\prime}$ are suppressed, which gives:
\begin{eqnarray}
    \beta(t_{g}) &=& -\frac{1}{\delta^{\prime}}\int^{t_{g}}_{0}dt^{\prime}\gamma^{2}(t^{\prime}) \nonumber \\
    \beta(t_{g})&=&-\frac{t_{g}\braket{\gamma^{2}}}{\delta^{\prime}},
\end{eqnarray}
where $\braket{\gamma^{2}}$ is the time averaged value of $\gamma^{2}(t)$. Every term in $\hat{U}_{2}$ is an integral similar to Eq.~(\ref{eq:beta_orig}). This, when considered with the arguments of Sec.~\ref{sec:u1}, allows us to analytically express the time propagator for the gate. For systems driven by DC ($\omega_{g}=0$) gradients this gives:
\begin{eqnarray}\label{eq:dc_prop}
    \hat{U}_{g} \simeq \exp\Big(i\braket{\gamma^{2}}t_{g}\sum_{j}\frac{\Omega_{j}^{2}}{\delta_{j}}\hat{S}_{z,j}^{2} \Big),
\end{eqnarray}
and for RF gradients ($\omega_{g}\neq 0$), we get:
\begin{eqnarray}\label{eq:rf_prop}\label{eq:final_ug}
    \hat{U}_{g} \simeq \exp\Big(\frac{i}{2}\braket{\gamma^{2}}t_{g}\sum_{j}\frac{\omega_{j}\Omega_{j}^{2}}{\omega_{j}^{2}-\omega_{g}^{2}}\hat{S}_{z,j}^{2} \Big).
\end{eqnarray} \\

\subsection*{Mixed-species crystals}
The results of this section apply to mixed-species as well as same-species crystals. This indicates AESE could be used to entangle two different ion species in a quantum logic spectroscopy experiment \cite{schmidt_2005}, or two same species ions in the presence of one or more sympathetic cooling ions. If we assume the sympathetic cooling ions are initialized to a specific internal state, Eq.~(\ref{eq:final_ug}) indicates their effect will be to add a $\propto~ \hat{\sigma}_{z,j}$ shift to each ion $j$. Since these are single qubit shifts that commute with $\hat{U}_{g}$, they can be tracked or coherently canceled with a spin-echo (a spin echo has the added benefit that it would remove the impact of spectator ions even in their state preparation is not perfect).

\section{Electronic gradients}\label{sec:electronic}

We now discuss how to implement electronically generated magnetic field gradient gates with AESE; since we can electronically ramp the gradient generating current on and off, we can directly shape $\gamma(t)$. For simplicity, we here set $\gamma(t)=\sin^{2}(\pi t/2\tau)$ and $\tau=t_{g}/2$ in the examples below, ramping the spin-dependent force for the entire gate duration; in many cases, however, AESE can be achieved when $\tau\ll t_{g}$. Assuming the magnetic field gradient strength $150$~T/m achieved in Ref.~\cite{srinivas_2021}, we examine a crystal of two $^{40}\mathrm{Ca}^{+}$ ions, each with a magnetic field sensitivity of $\frac{\partial\omega_{0}}{\partial B}/2\pi \simeq 2.5\times 10^{10}$ Hz/T. We also assume the gate operates on one pair of radial modes, where the center of mass (COM) mode is set to $\omega_{c}/2\pi=3~\mathrm{MHz}$, which has a collective Pauli operator $\hat{S}_{z,c}=\hat{\sigma}_{z,1}+\hat{\sigma}_{z,2}$ and a gradient Rabi frequency $\Omega_{c}$. We vary the radial stretch mode frequency $\omega_{s}$, which has a corresponding collective Pauli operator $\hat{S}_{z,s}=\hat{\sigma}_{z,1}-\hat{\sigma}_{z,2}$, and gradient Rabi frequency $\Omega_{s}$.  

\subsection{DC}\label{sec:dc}

For DC gradients, the effective detunings from the gating modes are the mode frequencies themselves, reducing Eq.~(\ref{eq:first_geo}) to:
\begin{eqnarray}\label{eq:ham_dc}
    \hat{H}_{\mathrm{DC}} = \hbar\gamma(t)\Big(\hat{S}_{z,c}\Omega_{c}[\hat{a}_{c}^{\dagger}e^{i\omega_{c}t} +\hat{a}_{c}e^{-i\omega_{c}t}]+\hat{S}_{z,s}\Omega_{s}[\hat{a}_{s}^{\dagger}e^{i\omega_{s}t} +\hat{a}_{s}e^{-i\omega_{s}t}]\Big) \nonumber \\
\end{eqnarray} 
We AESE the system by choosing parameters such that $2\pi/\omega_{\alpha}\ll \tau$. When this condition is met, we can write down the propagator for the system:
\begin{eqnarray}
    \hat{U}_{\mathrm{DC}} & \simeq &\exp\Big(i\braket{\gamma^{2}}t_{g}\Big[\frac{\Omega_{c}^{2}}{\omega_{c}}\hat{S}_{z,c}^{2} + \frac{\Omega_{s}^{2}}{\omega_{s}}\hat{S}_{z,s}^{2}\Big]\Big) \nonumber \\
    &=& \exp\Big(-i\Omega_{\mathrm{DC}}t_{g}\hat{\sigma}_{z,1}\hat{\sigma}_{z,2}\Big), \\
\end{eqnarray}
giving an effective gate frequency of:
\begin{eqnarray}
    \Omega_{\mathrm{DC}}\equiv \frac{3}{4}\Big(\frac{\Omega_{s}^{2}}{\omega_{s}}-\frac{\Omega_{c}^{2}}{\omega_{c}}\Big),
\end{eqnarray}
where we have substituted $\braket{\gamma^{2}}=3/8$. For this case, the entanglement generated by the COM mode coherently cancels the entanglement generated by the STR mode. Noting that for a given gradient $\Omega_{\alpha}^{2}~\propto ~ 1/\omega_{\alpha}$, which means $\Omega_{\mathrm{DC}}~\propto~ \omega_{s}^{-2}-\omega_{c}^{-2}$, we see that performing the DC version of the scheme in a regime where $\omega_{s}$ is at least a factor of two or three smaller than $\omega_{c}$ is desirable; for example, if $\omega_{s}/2\pi = 1~\mathrm{MHz}$, $\Omega_{\mathrm{DC}}$ is $8/9$ of what it would be in the absence of the COM mode. Further, the STR mode's symmetry typically makes its heating rate negligible compared to the COM mode \cite{harty_2016,srinivas_2021}, potentially allowing operation in low frequency regimes that have larger Lamb-Dicke factors, and adding another benefit to performing the gate in a regime where $\omega_{s}$ is small compared to $\omega_{c}$, i.e. near the ``zig-zag" instability.

\subsection{RF}\label{sec:rf}

When the gradient generating current oscillates at RF frequencies ($\omega_{g}\sim \mathrm{MHz}$) we can choose the gradient's detuning $\delta_{\alpha}\equiv \pm(\omega-\omega_{g})$ from at least one of the gating modes\textemdash a technique used in almost every geometric phase gate scheme. This allows us to enhance the gate speed and choose the primary gating mode we entangle the spins with, the latter being advantageous because it allows us to operate on modes that heat slowly due to symmetry (see Sec.~\ref{sec:heating} for more details). \\

\noindent The Hamiltonian for this system is given by:
\begin{eqnarray}\label{eq:ham_rf}
    \hat{H}_{\mathrm{RF}} = \hbar\gamma(t)\cos(\omega_{g}t)\Big(\hat{S}_{z,c}\Omega_{c}[\hat{a}_{c}^{\dagger}e^{i\omega_{c}t} +\hat{a}_{c}e^{-i\omega_{c}t}]+\hat{S}_{z,s}\Omega_{s}[\hat{a}_{s}^{\dagger}e^{i\omega_{s}t} +\hat{a}_{s}e^{-i\omega_{s}t}]\Big), \nonumber \\
\end{eqnarray} 
which gives an effective gate speed of:
\begin{eqnarray}
    \Omega_{\mathrm{RF}} = \frac{3}{16}\Big|\frac{\omega_{s}\Omega_{s}^{2}}{\omega_{s}^{2}-\omega_{g}^{2}} - \frac{\omega_{c}\Omega_{c}^{2}}{\omega_{c}^{2}-\omega_{g}^{2}}\Big|.
\end{eqnarray}
While splitting the gradient into $\pm\omega_{g}$ frequency components ostensibly results in an immediate factor of $4$ reduction in speed, we can more than compensate for this reduction by tuning $\omega_{g}$ such that $\omega_{s}/|\omega_{s}^{2}-\omega_{g}^{2}|\gg 1/\omega_{s}$. Further, we can also operate with an $\omega_{g}$ that is blue detuned from $\omega_{s}$, i.e. $\omega_{g}>\omega_{s}$. The ability to control the sign of $\delta_{\alpha}$ allows us to ensure the entanglement generated by each mode coherently adds instead of cancels. Therefore, using an RF gradient not only lets us further enhance the speed of the gate, but also relax the restriction that we operate in a regime where the gating mode frequencies are significantly split; this reduces the speed penalty for operating in a regime where $\omega_{s}\sim \omega_{c}$. \\

\begin{figure}[b]
\includegraphics[width=1.1\textwidth]{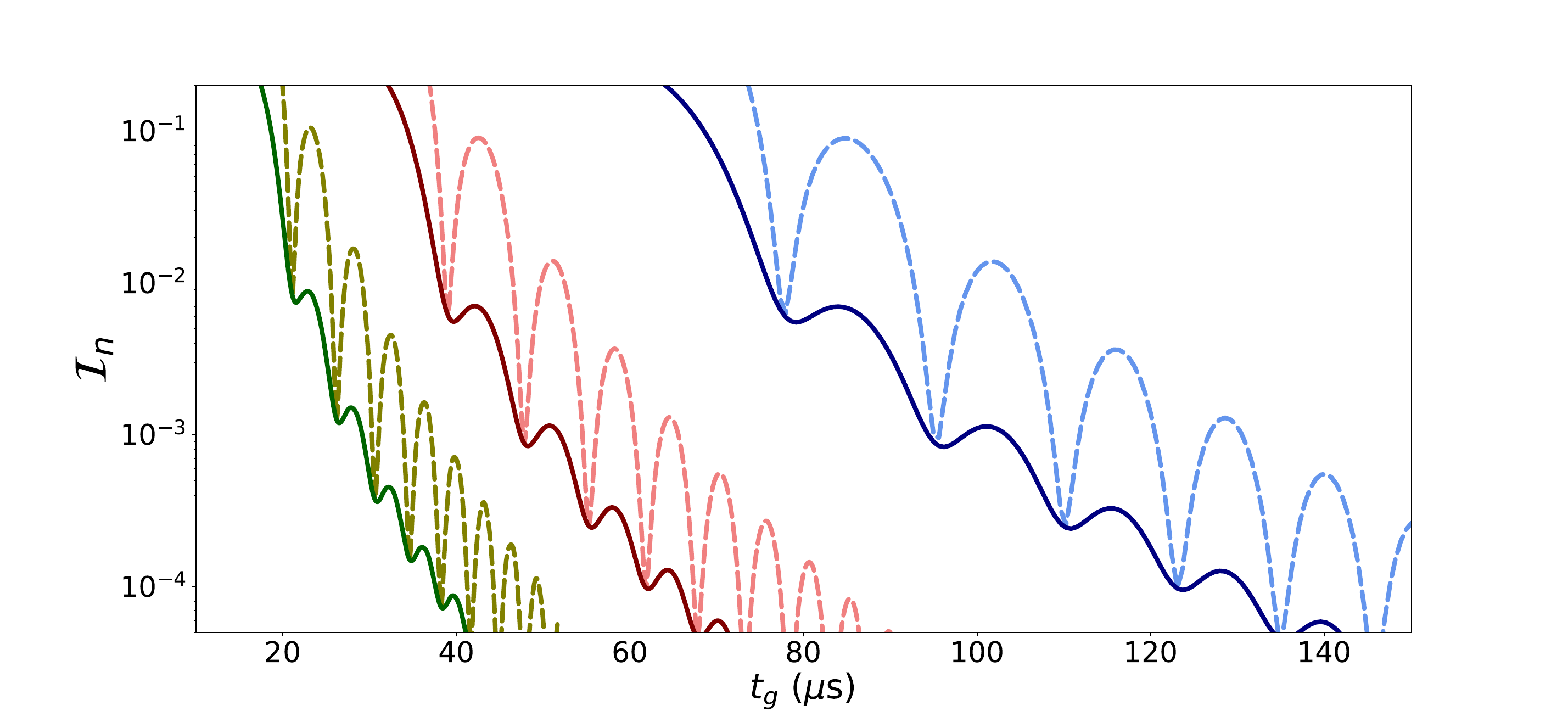}
\centering
\caption{Infidelity $\mathcal{I}_{n}$ versus gate time $t_{g}$ averaged over SU(4). Here the spin-motion coupling is adiabatically eliminated by having the gradient follow a $\gamma(t)=\sin^2({\pi t/t_{g}}$) envelope. In each calculation, we set $\omega_{c}/2\pi = 3~\mathrm{MHz}$ and show results for states initialized to $n=0$ (solid) and $n=10$ (dashed), representing $\mathcal{I}_{n}$ for ground state and Doppler cooled crystals, respectively. We show this for a crystal of two $^{40}\mathrm{Ca}^{+}$ ions driven by a $300~\mathrm{T/m}$ gradient when $\omega_{s}/2\pi = 250~\mathrm{kHz}$ (green bottom), $300~\mathrm{T/m}$ and $\omega_{s}/2\pi = 1~\mathrm{MHz}$ (red middle), as well as for $150~\mathrm{T/m}$ and $\omega_{s}/2\pi = 1~\mathrm{MHz}$ (blue top).}
\label{fig:rf_tau_scan}
\end{figure}

\noindent In Fig.~\ref{fig:rf_tau_scan}, we provide numerical calculations of the of the RF version of this scheme's infidelity, showing how AESE eliminates errors from residual spin-motion entanglement\textemdash even without ground-state cooling\textemdash for realistic experimental parameters. In each example, we pulse shape over the entire interaction, setting $\gamma(t)=\sin^{2}(\pi t/t_{g})$ (i.e. $\tau=t_{g}/2$). This allows us to minimize the value of $\delta_{s}$ needed to AESE for a given $t_{g}$. For a target qubit state $\ket{T}$, the infidelity is:
\begin{eqnarray}
    \mathcal{I}_{n} = 1-\sum_{n^{\prime}}|\bra{T}\braket{n^{\prime}|\psi(t_{g})}\ket{n}|^{2},
\end{eqnarray}
where we are tracing over the motion, and assuming a system initialized to $\ket{\psi(0)}\ket{n}$. We here increase the value of $t_{g}$ by increasing the value of $\omega_{g}$. Increasing $\omega_{g}$ increases the magnitude of $\delta_{s}$ and $\tau$, both of which parametrically suppress the residual spin-motion entanglement; this indicates a clear trade-off between gate speed and motional insensitivity, the optimal parameters likely depending on the specific situation. In Fig.~\ref{fig:rf_tau_scan} we take the relevant gating modes to be the radial COM $\omega_{c}/2\pi=3~\mathrm{MHz}$, and the radial STR mode, where $\omega_{s}/2\pi = 250~\mathrm{kHz}$ (green left plot) and $\omega_{s}/2\pi = 1~\mathrm{MHz}$ (red center and blue right plots). The green and red plots use the same gradient strength, differing only in the stretch mode frequency, highlighting that one can achieve a given error with a faster gate by operating close to the zig-zag instability (due to the $\propto~\omega_{s}^{-1/2}$ dependence of $\Omega_{s}$). We also show the dependence of $\mathcal{I}_{n}(t_{g})$ on the size of the magnetic field gradient, using both $300~\mathrm{T/m}$ (green left, red center) and $150~\mathrm{T/m}$ (blue right). Further, we calculate $\mathcal{I}_{n}$ for systems initialized to the ground state of motion $n=0$ (solid), and the state where $n=10$ (dashed) to model the gate when the crystal is ground state cooled and at Doppler temperatures, respectively. We can see that both $n=0$ and $n=10$ periodically converge to one another when $\hat{U}_{1}\rightarrow \hat{I}$, while both plots continue to improve with $t_{g}$ as the off-resonant terms in $\hat{U}_{2}$ are AESEd. Fig.~\ref{fig:rf_tau_scan} demonstrates that AESE should enable high fidelity gates even at the Doppler temperature, and while achieving relatively small values of $t_{g}$.

\section{Ferromagnetic gradients}\label{sec:ferromagnets}

It is also possible  to generate magnetic field gradients with permanent magnets \cite{mintert_2001,weidt_2016,arrazola_2018,barthel_2022}. While electronically induced gradients have, so far, resulted in faster, higher fidelity two-qubit gates \cite{srinivas_2021}, both the resistive heating from the gradient-generating currents and the need for additional control signals pose potential scaling bottlenecks. For this reason, it could be desirable to perform gates with AESE using ferromagnetic gradients. Since we cannot turn a permanent magnet on/off, however, we cannot modulate the spin-dependent force directly, necessitating some alternative technique for controlling its strength. To this end, we propose \textit{effectively} ramping the spin-motion coupling via shelving to/from from a `clock' transition, where information is stored in a pair of states with a frequency difference that has no linear dependence on magnetic field strength, to a `Zeeman' transition, where information is stored in a pair of states with a frequency difference that is highly-dependent on magnetic field strength. Here, we consider a three state system comprising a clock state $\ket{1\mathrm{c}}$, a Zeeman state $\ket{1\mathrm{z}}$, and a ground state $\ket{0}$. An example of such a system is the $^{2}\mathrm{S}_{1/2}$ manifold of $^{171}\mathrm{Yb}^{+}$, where $\ket{0}\equiv\ket{F=0,m_{f}=0}$, $\ket{1\mathrm{c}}\equiv \ket{F=1,m_{f}=0}$, and $\ket{1\mathrm{z}}\equiv \ket{1\mathrm{z}}\equiv\ket{F=1,m_{f}=1}$. For simplicity, we only consider one mode in this section, but note the requirement of operating in a regime where $\omega_{\mathrm{s}}$ is considerably less than $\omega_{\mathrm{c}}$ still holds. \\

\noindent Assuming the ions only experience a spin-dependent force when in the $\ket{1\mathrm{z}}$ state, the Hamiltonian is
\begin{eqnarray}\label{eq:proj_z_gate}
    \hat{H}_{\mathrm{z}} = 2\hbar\Omega_{\mathrm{z}}\hat{P}_{\mathrm{z}}(\hat{a}^{\dagger}e^{i\omega t} + \hat{a}e^{-i\omega t}),
\end{eqnarray}
where $\hat{P}_{z}\equiv \ket{1\mathrm{z}}_{1}\bra{1\mathrm{z}} - \ket{1\mathrm{z}}_{2}\bra{1\mathrm{z}}$ is the collective projection operator for state $\ket{1\mathrm{z}}$ acting on one of the radial STR modes. Note that there is an added factor of $2$ in Eq.~(\ref{eq:proj_z_gate}) because we are representing the system's spin-dependence with a projection onto $\ket{1\mathrm{z}}$ rather than its frequency difference with the $\ket{0}$ state. In the following, we describe two schemes for generating the single-qubit transformation:
\begin{eqnarray}\label{eq:shelve_u}
    \hat{U}_{\mathrm{sh}} = \prod_{j}\exp\Big(-i\phi\hat{\sigma}_{x,j}^{\mathrm{cz}}\Big),
\end{eqnarray}
where $\sigma_{x,j}^{\mathrm{cz}}$ is a Pauli-x operator acting on qubit $j$ in the $\{\ket{1\mathrm{z}},\ket{1\mathrm{c}}\}$ subspace. Transforming into a rotating frame with respect to this operator acting on both qubit ions, $\hat{U}_{\mathrm{sh}}^{\dagger}\hat{H}_{\mathrm{z}}\hat{U}_{\mathrm{sh}}$, gives:
\begin{eqnarray}\label{eq:interaction}
    \tilde{H}_{\mathrm{z}}(t) &=& 2\hbar\Omega_{\mathrm{z}}\Big(\hat{P}_{\mathrm{z}}\cos^{2}[\phi] + \hat{P}_{\mathrm{c}}\sin^{2}[\phi]+\frac{1}{2}\sin[2\phi]\hat{S}_{y}^{\mathrm{cz}} \Big)\Big(\hat{a}^{\dagger}e^{i\omega t} + \hat{a}e^{-i\omega t}\Big), \nonumber \\
\end{eqnarray}
where we have introduced $\hat{S}_{x}^{\mathrm{cz}}\equiv \sum_{j}\ket{1\mathrm{z}}_{j}\bra{1\mathrm{c}} + \ket{1\mathrm{c}}_{j}\bra{1\mathrm{z}}$, $\hat{S}_{y}^{\mathrm{cz}}\equiv \sum_{j}i(\ket{1\mathrm{z}}_{j}\bra{1\mathrm{c}} - \ket{1\mathrm{c}}_{j}\bra{1\mathrm{z}}$) and $\hat{S}_{z}^{\mathrm{cz}}\equiv \sum_{j}\ket{1\mathrm{c}}_{j}\bra{1\mathrm{c}} - \ket{1\mathrm{z}}_{j}\bra{1\mathrm{z}}$, which correspond to collective operators acting on the $\{\ket{1\mathrm{z}},\ket{1\mathrm{c}}\}$ subspace of the qubits. When we generate Eq.~(\ref{eq:interaction}) using the `Rabi flopping' scheme described below, we exactly generate this transformation. When we use the `Adiabatic Rapid Passage' scheme, however, we generate Eq.~(\ref{eq:interaction}) with an additional dressed state splitting term (assuming the operation is adiabatic). Assuming the states of the system are initialized to the $\{\ket{0},\ket{1\mathrm{c}}\}$ subspace, we can see that $\tilde{H}_{\mathrm{z}}(0)\ket{\psi(0)}=0$, i.e. there is no spin-motion coupling. During a shelving time $\tau$, the value of $\phi$ is first smoothly increased to $\pi/2$, reaching the maximum strength of the spin-motion coupling, then smoothly to $0$ or up to $\pi$, ramping the coupling back down to zero. As we will show, this smooth increase/decrease of the \textit{effective} spin-motion coupling AESEs the system even though the magnetic field gradient cannot be controlled itself. \\

\noindent This method, unfortunately, generates an intrinsic source of error not present in gates that use electronic gradients. This is due to the MS-like:
\begin{eqnarray}
    \hat{H}_{\mathrm{e}}\equiv \hbar\Omega_{z}\sin(2\phi)\hat{S}_{y}^{\mathrm{cz}}(\hat{a}^{\dagger}e^{i\omega t}+\hat{a}e^{-i\omega t}) 
\end{eqnarray}
error Hamiltonian in Eq.~(\ref{eq:interaction}) generating unwanted entanglement between the motion and the $\ket{1\mathrm{z}}\leftrightarrow\ket{1\mathrm{c}}$ transition. This effect could limit gate fidelity for faster gates, and when larger values of $\tau$ are needed to AESE the system\textemdash if implemented with a Rabi transition. When using adiabatic rapid passage ARP, however, we this effect can be suppressed arbitrarily via increasing the dressed-state splitting of the transition. 

\subsubsection*{Rabi flopping}

\begin{figure}[b]
\includegraphics[width=1.1\textwidth]{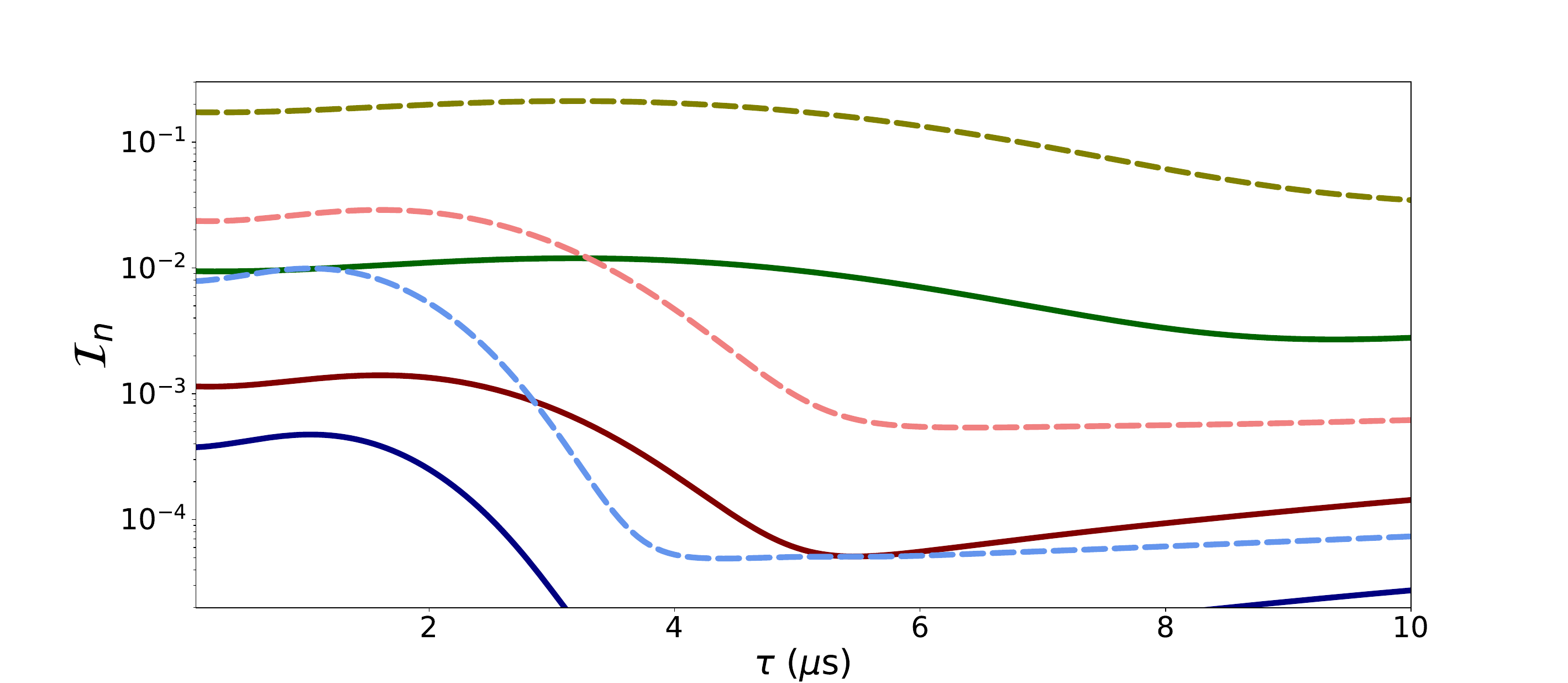}
\centering
\caption{Infidelity $\mathcal{I}_{n}$ versus Rabi shelving/ramp time $\tau$ from a magnetic field insensitive `clock' state to a magnetic field sensitive `Zeeman' state upon averaging over SU(4). We choose a gradient of $50~\mathrm{T/m}$ and set $\omega/2\pi = 250~\mathrm{kHz}$ (green), $\omega/2\pi = 500~\mathrm{kHz}$ (red), and $\omega/2\pi = 750~\mathrm{kHz}$ (blue), showing results for states initialized to $n=0$ (solid) and $n=10$ (dashed).}
\label{fig:rabi_scan}
\end{figure}

We now consider a system evolving under Eq.~(\ref{eq:proj_z_gate}) with an additional Rabi transition term, which will map the system between the $\ket{1\mathrm{z}}$ and $\ket{1\mathrm{c}}$ states; one can do this with either lasers or microwaves. The full gating Hamiltonian is then: 
\begin{eqnarray}\label{eq:rabi_gate}
    \hat{H}_{\mathrm{r}} &=& 2\hbar\Omega_{\mathrm{z}}\hat{P}_{\mathrm{z}}(\hat{a}^{\dagger}e^{i\omega t} + \hat{a}e^{-i\omega t}) + \frac{\hbar\Omega_{\mathrm{R}}(t)}{2}\hat{S}_{x}^{\mathrm{cz}},
\end{eqnarray}
where we have introduced $\Omega_{\mathrm{R}}(t)$ to represent the time-dependent Rabi frequency. Here, $\Omega_{\mathrm{R}}(t)$ is constrained such that it is only non-zero during the shelving times $\tau$, and that $\int^{t_{0}+\tau}_{t_{0}}dt^{\prime}\Omega_{\mathrm{R}}(t^{\prime})=\pm\pi$ for each shelving period. Transforming into the interaction picture with respect to this term gives Eq.~(\ref{eq:interaction}) where:
\begin{eqnarray}
    \phi(t) = \frac{1}{2}\int^{t}_{0}dt^{\prime}\Omega_{\mathrm{R}}(t^{\prime}),
\end{eqnarray}
showing the desired ramping of the \textit{effective} spin-motion coupling of the system. If we choose $\Omega_{\mathrm{R}}(t)$ such that it has the opposite sign when shelving from $\ket{1\mathrm{z}}\rightarrow\ket{1\mathrm{c}}$ compared to when shelving from $\ket{1\mathrm{z}}\leftarrow\ket{1\mathrm{c}}$, then $\phi(t_{g})=0$, and the state of the interaction picture wavefunction corresponds exactly to that of the lab frame wavefunction \cite{sutherland_2019}. Therefore, we can safely think of the dynamics of the system as resulting from Eq.~(\ref{eq:interaction}). If we assume the initial state is within the clock state manifold and ignore the term $\propto~ \hat{S}_{y}^{\rm cz}$, this Hamiltonian implements an entangling gate with AESE. \\

\noindent In Fig.~\ref{fig:rabi_scan} we explore the efficacy of this procedure for a sequence such that $\Omega_{\mathrm{R}}(t)=2\pi \sin^{2}(\pi t/\tau)/\tau$ when $t<\tau$ and $\Omega_{\mathrm{R}}(t)=-2\pi \sin^{2}(\pi t/\tau)/\tau$ when $t>t_{g}-\tau$, including the effects of $H_{\rm e}$. We choose a smooth envelope (as opposed to a square one) for $\Omega_{\mathrm{R}}(t)$ in order to suppress off-resonant $1^{\mathrm{st}}$ order effects from $\hat{H}_{\mathrm{e}}$ by ensuring $\dot{\hat{H}}_{\mathrm{e}}(0)=\dot{\hat{H}}_{\mathrm{e}}(\tau)=\dot{\hat{H}}_{\mathrm{e}}(t_{g}-\tau)=\dot{\hat{H}}_{\mathrm{e}}(t_{g})=0$. In the figure, we simulate chains of two $^{40}\mathrm{Ca}^{+}$ ions experiencing a $50~\mathrm{T/m}$ gradient. We show gate infidelity versus $\tau$ for mode frequencies of $\omega_{s}/2\pi=250~\mathrm{kHz}$ ($t_{g}\simeq 170~\mu\mathrm{s}$), $\omega/2\pi=500~\mathrm{kHz}$ ($t_{g}\simeq 650~\mu\mathrm{s}$), and $\omega/2\pi=1~\mathrm{MHz}$ ($t_{g}\simeq 2.5~\mathrm{ms}$). In each plot, we can see that, initially, increasing $\tau$ decreases $\mathcal{I}_{n}$ because we are reaching the limit $\omega\gg 1/\tau$ and achieving AESE. For larger values of $\tau$, however, we see that $\mathcal{I}_{n}$ begins to increase simply because the system is exposed to $\hat{H}_{\mathrm{e}}$ for longer. Further, we see that choosing larger values of $\omega$ decreases the minimum achievable value of $\mathcal{I}_{n}$ because the motional component of $\hat{H}_{\mathrm{e}}$ becomes more and more off-resonant. This indicates that, for a given gradient and value of $\omega$, we can suppress the infidelity caused by $\hat{H}_{\mathrm{e}}$ by increasing $\omega$, and, unfortunately, the value of $t_{g}$. In the next section, we discuss a way of suppressing the effects of $\hat{H}_{\mathrm{e}}$ without sacrificing gate speed.

\subsubsection*{Adiabatic rapid passage}

We can also induce $\ket{1\mathrm{z}}\leftrightarrow\ket{1\mathrm{c}}$ shelving with an adiabatic rapid passage ARP transition. To show this, we consider Eq.~(\ref{eq:proj_z_gate}) with an additional set of single-qubit interactions:
\begin{eqnarray}\label{eq:arp_gate}
    \hat{H}_{\mathrm{arp}} &=& 2\hbar\Omega_{\mathrm{z}}\hat{P}_{\mathrm{z}}(\hat{a}^{\dagger}e^{i\omega t} + \hat{a}e^{-i\omega t}) + \frac{\hbar\Omega_{\mathrm{y}}(t)}{2}\hat{S}_{y}^{\mathrm{cz}} + \frac{\hbar\Delta(t)}{2}\hat{S}_{z}^{\mathrm{cz}}.
\end{eqnarray}
The $\hat{U}_{\mathrm{sh}}$ operator given by Eq.~(\ref{eq:shelve_u}) diagonalizes these added terms when $\phi = \frac{1}{2}\arctan[\Omega_{\mathrm{y}}(t)/\Delta(t)]$. Transforming $\hat{H}_{\mathrm{arp}}$ in this way gives:
\begin{eqnarray}
    \tilde{H}_{\mathrm{arp}} &=& \hat{U}^{\dagger}_{\mathrm{sh}}(t)\hat{H}_{\mathrm{arp}}\hat{U}_{\mathrm{sh}}(t) + i\hbar\dot{\hat{U}}^{\dagger}_{\mathrm{sh}}(t)\hat{U}_{\mathrm{sh}}(t) \nonumber \\
    &=& \tilde{H}_{\mathrm{z}} + \frac{\Delta_{\mathrm{d}}(t)}{2}\hat{S}_{\mathrm{z}}^{\mathrm{cz}} - \hbar\dot{\phi}\hat{S}_{x}^{\mathrm{cz}},
\end{eqnarray}
where $\Delta_{\mathrm{d}}(t)\equiv [\Omega_{\mathrm{y}}^{2}(t)+\Delta^{2}(t)]^{1/2}$ is the splitting of the $\ket{\pm}$ dressed states defined by $\hat{U}_{\mathrm{sh}}$. To see how we can use this to shelve, consider a system initialized to the $\{\ket{1\mathrm{c}},\ket{0}\}$ manifold at a time $t_{0}$, with boundary conditions: $\Delta(t_{0})= -\Delta(t_{0}+\tau)=\Delta_{0}$, $\Omega_{\mathrm{y}}(t_{0})=\Omega_{\mathrm{y}}(t_{0}+\tau)=\Delta(t_{0}+\tau/2)=0$, and $\Omega_{\mathrm{y}}(t_{0}+\tau/2)=\pm\Delta_{0}$. If we choose functional forms of $\Omega_{\mathrm{y}}(t)$ and $\Delta(t)$ that smoothly transition between these boundary conditions, we can induce the $\ket{1\mathrm{c}}\leftrightarrow\ket{1\mathrm{z}}$ shelving we want; then, if we set $\tau\gg 2\pi /\omega$, we also perform the gate with AESE. We can see that the additional $\propto~\hat{S}^{\mathrm{cz}}_{z}$ commutes with every term in Eq.~(\ref{eq:proj_z_gate}) except the $\propto~\hat{S}_{y}^{\mathrm{cz}}$ component of $\hat{H}_{\mathrm{e}}$ and the $\propto~\dot{\phi}(t)$ diabatic term, both of which are rotated in the $\mathrm{xy}-$plane of the Bloch sphere at a frequency $\Delta_{\mathrm{d}}$. In the limit $\Delta_{\mathrm{d}}(t)\gg \Omega_{\mathrm{z}}$, the effect of $\hat{H}_{\mathrm{e}}$ averages to zero. This allows us to increase $\tau$ to AESE the gate, removing infidelities from residual spin-motion coupling, without increasing the infidelity due to $\hat{H}_{\mathrm{e}}$. At the same time $\dot{\phi}(t)~\propto~ 1/\tau$, meaning diabaticity's effect on the gate fidelity is also suppressed for large values of $\tau$. \\
\begin{figure}[b]
\includegraphics[width=1.1\textwidth]{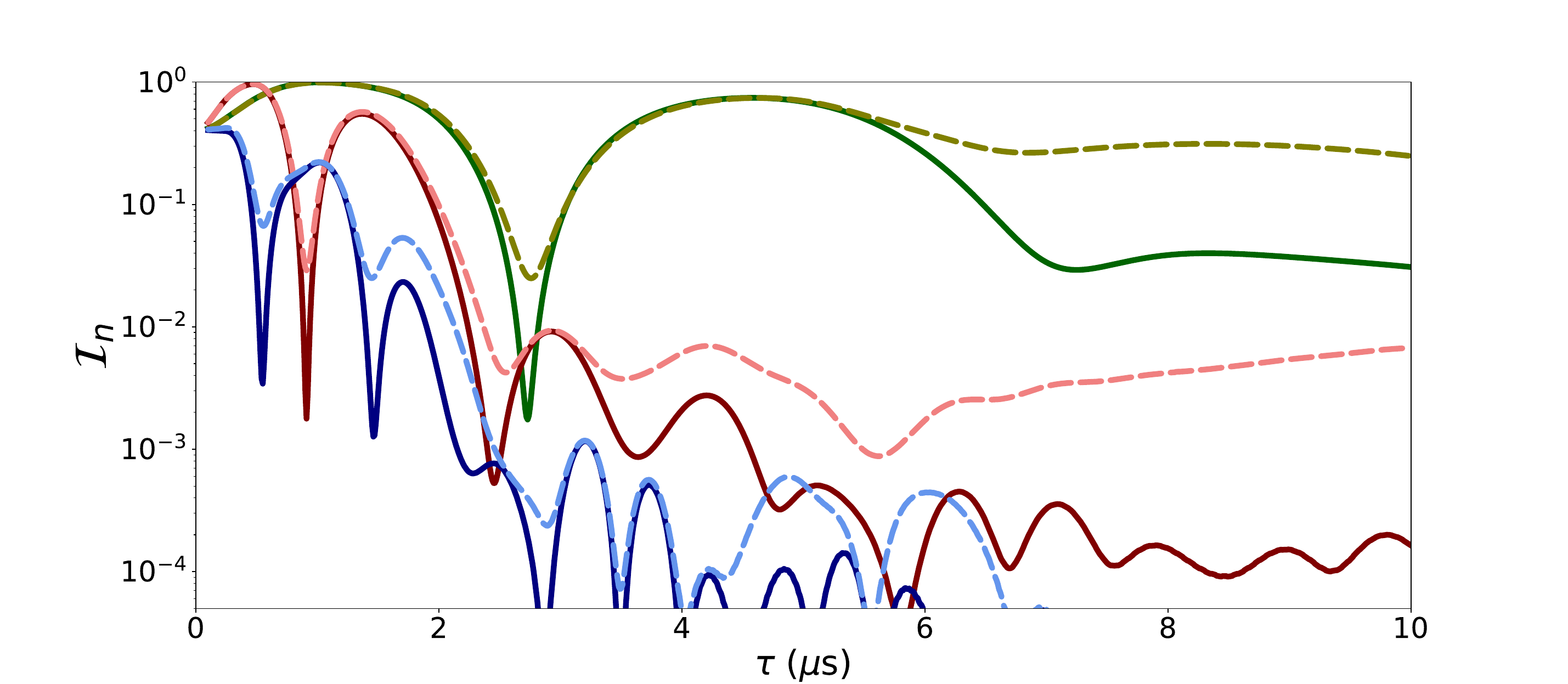}
\centering
\caption{Gate infidelity $\mathcal{I}_{n}$ versus shelving/ramp time $\tau$. Here we adiabatically eliminate the spin-motion coupling with an adiabatic rapid passage ARP transition from a magnetic field insensitive `clock' state to a magnetic field sensitive `Zeeman' state. We choose a gradient of $50~\mathrm{T/m}$ and set $\omega_{s}/2\pi = 500~\mathrm{kHz}$, showing results for states initialized to $n=0$ (solid) and $n=10$ (dashed). We show this for maximum dressed state splittings, $\Delta_{0}/2\pi$ in the text, of $400~\mathrm{kHz}$ (green top), $1.2~\mathrm{MHz}$ (red middle), and $2~\mathrm{MHz}$ (blue bottom).}
\label{fig:arp_scan}
\end{figure}

\noindent In Fig.~\ref{fig:arp_scan} we show $\mathcal{I}_{n}$ versus $\tau$ for a chain of two $^{40}\mathrm{Ca}^{+}$ ions coupled to their motion with a $50~\mathrm{T/m}$ gradient. We choose a gating mode frequency of $\omega/2\pi=250~\mathrm{kHz}$ for systems initialized to $n=0$ (solid lines) and $n=10$ (dashed lines). For each run, we set $\Delta(t^{\prime})=\Delta_{0}\cos(\pi t^{\prime}/\tau)$ and $\Omega_{y}(t^{\prime})=\Delta_{0}\sin^{2}(\pi t^{\prime}/\tau)$, where $t^{\prime}=t$ when $t<\tau$, $t^{\prime}=\tau$ when $\tau\leq
t \leq t_{g}-\tau$, and $t^{\prime}=t+2\tau-t_{g}$ when $t>t_{g}-\tau$; this makes the dressed state splitting $\Delta_{\mathrm{d}}(t)=\Delta_{0}[\cos^{2}(\pi t^{\prime}/\tau)+\sin^{4}(\pi t^{\prime}/\tau)]^{1/2}$. We chose these functions for $\Delta(t)$ and $\Omega_{\mathrm{y}}(t)$ because they meet the boundary conditions described above, while ensuring $\Delta(t)$ and $\Omega_{\mathrm{y}}(t)$ vanish at the end of each shelving sequence to suppress unwanted off-resonant effects. While we could also suppress effect of $\hat{H}_{\mathrm{e}}$ by increasing $\omega$ and slowing the gate, we can now decrease the infidelities due to $\hat{H}_{\mathrm{e}}$ by increasing $\Delta_{\mathrm{d}}$, which does not slow the gate. Further, we can now, in theory, suppress the infidelity from $\hat{H}_{\mathrm{e}}$ arbitrarily by continuing to increasing the shelving fields strengths, and thus the values of $\Delta_{d}(t)$, during the sequence. 

\section{Motional errors}

Arguably the biggest benefit to performing geometric phase gates with AESE is their insensitivity to motional errors, which typically comprise a significant portion of laser-free two-qubit gates'  error budgets \cite{harty_2016,weidt_2016, zarantonello_2019,srinivas_2021}. While it is possible to modify the phase space trajectory or add dynamical decoupling sequences that decrease a gate's sensitivity to certain sources of motional errors without AESE \cite{hayes_2012, haddadfarshi_2016,webb_2018, shapira_2018,zarantonello_2019,sutherland_2020}, these techniques have (so far) offered marginal improvements to their respective motional error budgets. In this section, we discuss how performing geometric phase gates with AESE significantly decreases a gate's sensitivity to motional noise relative to other laser-free schemes. We also discuss how AESE eases the need for ground state cooling to reach high fidelities. We ignore the counter-rotating terms in the RF gradient case, and introduce $\delta_{s(c)}\equiv \omega_{g}-\omega_{s(c)}$ to represent the mode detunings and $\Omega_{j}^{\prime}=\Omega_{j}(\Omega_{j}/2)$ for DC(RF) gradients in order to keep the results general to both.

\subsection{Heating}\label{sec:heating}

We now discuss the infidelities from heating when performing a two-qubit gate with AESE. By `heating', we here mean motional excitations due to a bath of extraneous electric fields that perturb the crystal. When qubits experience a spin-dependent force, the displacements from this E-field bath induce a shift proportional to the spin component of the gate \cite{sutherland_2022_1}. While we are proposing a new parameter regime, this scheme is still a geometric phase gate\cite{molmer_1999,molmer_2000,leibfried_2002,roos_2008,steane_2014} and previous derivations of heating-induced infidelities apply. Following the prescription outlined in Ref.~\cite{sutherland_2022_1}, we model this effect with a time-dependent E-field oscillating at frequency $\omega_{e}$, which gives:
\begin{eqnarray}\label{eq:ham_added_e_field}
\hat{H}_{g}=\hbar\gamma(t)\sum_{\alpha=s,c}\Big(\hat{S}_{z,\alpha}\Omega_{\alpha}^{\prime}[\hat{a}_{\alpha}^{\dagger}e^{i\delta_{\alpha}t} +\hat{a}_{\alpha}e^{-i\delta_{\alpha}t}] + 2\hbar \cos(\omega_{e}t) g_{\alpha}[\hat{a}_{\alpha}^{\dagger}e^{i\omega_{\alpha}t} +\hat{a}_{\alpha}e^{-i\omega_{\alpha}t}]\Big), \nonumber \\
\end{eqnarray}
where $g_{\alpha}$ is the Rabi frequency of the E-field projected onto mode $\alpha$. Following Ref.~\cite{sutherland_2022_1}, we transform into the interaction picture with respect to the E-field terms, giving:
\begin{eqnarray}\label{eq:heat_interaction}
\hat{H}_{g,I}=\hbar\gamma(t)\sum_{\alpha=s,c}\hat{S}_{z,\alpha}\Omega_{\alpha}^{\prime}[(\hat{a}_{\alpha}^{\dagger}+d_{\alpha}^{*})e^{i\delta_{\alpha}t} +(\hat{a}_{\alpha}+d_{\alpha})e^{-i\delta_{\alpha}t}],
\end{eqnarray}
where:
\begin{eqnarray}
    d_{\alpha}\equiv g_{\alpha}\Big(\frac{1-e^{i[\omega_{\alpha}+\omega_{e}]t}}{\omega_{\alpha}+\omega_{e}} +  \frac{1-e^{i[\omega_{\alpha}-\omega_{e}]t}}{\omega_{\alpha}-\omega_{e}}\Big),
\end{eqnarray} 
is the displacement due to the E-field force. There are two spectral regimes where E-field noise has the potential to cause significant errors: when $2\pi/\omega_{e} \gtrsim t_{g}$ and when $\omega_{e}\sim \omega_{\alpha}$, for any $\alpha$. \\ 

\subsubsection*{Low-frequency E-field noise}\label{sec:lf_noise}
For a DC gradient, i.e. when $\delta_{c(s)}=\omega_{c(s)}$, $\omega_{g}=0$ , and $\Omega_{c(s)}^{\prime}=\Omega_{c(s)}$, Eq.~(\ref{eq:heat_interaction}) has a near-resonant term when $\omega_{e}\ll \Omega_{\alpha}^{\prime}$, giving a significant contribution from low-frequency LF E-field noise. Using this to drop counter-rotating terms in Eq.~(\ref{eq:heat_interaction}) gives:
\begin{eqnarray}\label{eq:lf_error_ham}
\hat{H}_{LF}=\hbar\gamma(t)\sum_{\alpha=s,c}\hat{S}_{z,\alpha}\Omega_{\alpha}[\hat{a}_{\alpha}^{\dagger}e^{i\omega_{\alpha}}t +\hat{a}_{\alpha}e^{-i\omega_{\alpha}t}] - \frac{4\hbar\Omega_{\alpha}g_{\alpha}\omega_{\alpha}}{\omega_{\alpha}^{2}-\omega_{e}^{2}}\cos(\omega_{e}t)\hat{S}_{z,\alpha}. \nonumber \\
\end{eqnarray}
Since E-field noise tends to decrease with $\omega_{e}$ \cite{blatt_2008}, sensitivity to small $\omega_{e}$ could be a significant source of infidelity for DC gradient schemes\textemdash if we do not contain the damage. Fortunately, the error term in Eq.~(\ref{eq:lf_error_ham}) commutes with the gate at all times, and the resulting infidelity can be strongly decreased with dynamical decoupling sequences \cite{hayes_2012}. If we use the RF gradient explored scheme explored in Sec.~\ref{sec:rf}, however, the error term in Eq.~(\ref{eq:lf_error_ham}) no longer contains a near-resonant term at $\omega_{e}\simeq 0$, making the gate insensitive to low-frequency E-field noise. \\

\subsubsection*{White noise heating}\label{eq:white_noise}
One of the main advantages of using the RF gradient scheme (discussed in Sec.~\ref{sec:rf}) is that the effect of low-frequency E-field noise is averaged to zero. If this is the case, the main effect of heating on the gate's error budget is from `Markovian' E-field noise near the gating mode frequencies, i.e. $\omega_{e}\sim \omega_{\alpha}$, which we assume to be a white noise bath. Reference~\cite{sutherland_2022_1} derived an analytic formula for the effect of this on a single-mode two-qubit gate. Here, the infidelity due to heating of multiple modes will be additive because the E-field terms in Eq.~(\ref{eq:ham_added_e_field}) commute when $\alpha \neq \alpha^{\prime}$. If a system has multiple modes, this gives:
\begin{eqnarray}
    \mathcal{I}_{\mathrm{h}} &=& 4t_{g}\sum_{\alpha}\frac{\dot{n}_{\alpha}\Omega_{\alpha}^{\prime 2}}{\delta_{\alpha}^{2}}\lambda_{\hat{S}_{z,\alpha}}^{2}\nonumber \\
    &=&\frac{16 t_{g}}{3}\sum_{\alpha}\frac{\dot{n}_{\alpha}\Omega_{\alpha}^{\prime 2}}{\delta_{\alpha}^{2}},
\end{eqnarray}
where, in the second line, we have averaged over all initial states, giving $\lambda_{\hat{S}_{z,s(c)}}^{2}=8/5$. This equation indicates another advantage of operating when tuned significantly closer to the STR mode than the COM mode. For same-species two-ion crystals, typically $\dot{n}_{s}\ll\dot{n}_{c}$ \cite{noguchi_2014, harty_2016,srinivas_2021} due to the fact that an external E-field couples to the COM mode directly but the STR differentially; this usually leads the STR mode to heat well over an order of magnitude more slowly than the COM mode. Using the numbers from the middle red plot in Fig.~\ref{fig:rf_tau_scan}, i.e. two $^{40}\mathrm{Ca}^{+}$ ions being driven by a $150~\mathrm{T/m}$ RF gradient, where $\omega_{s}/2\pi=1~\mathrm{MHz}$ and $\omega_{c}/2\pi=3~\mathrm{MHz}$, we can estimate $\mathcal{I}_{\mathrm{h}}$ in near-term experiments. Choosing $\omega_{g}/2\pi = 1.1~\mathrm{MHz}$ makes $\delta_{s}/2\pi = 100~\mathrm{kHz}$ and $\delta_{c}/2\pi = 1.9~\mathrm{MHz}$. Thus if we assume heating rates of $\dot{n}_{s}=1$ quanta/s and $\dot{n}_{c}=100$ quanta/s, we get $\mathcal{I}_{\mathrm{h}}\simeq 1\times 10^{-5}$.  \\

\subsection*{Mode frequency fluctuations}\label{sec:motional_decoherence}

If a gate's mode frequencies fluctuate during operation its phase space trajectory distorts, causing two distinct error-inducing mechanisms. The first is from residual spin-motion entanglement due to the system's phase space trajectory not closing. The infidelity from this mechanism scales linearly with initial phonon number, making it sensitive to temperature \cite{sutherland_2022_1}. The second mechanism is the ion chain's phase space trajectory encompassing the wrong area, which results in the wrong entanglement angle for the gate. The infidelity from this error mechanism does not scale with initial phonon number, meaning its effect on gate infidelity is independent of temperature \cite{sutherland_2022_1}. Here, we show that AESEing can suppress the former, \textit{temperature dependent} mechanism arbitrarily by increasing the values of $\delta_{\alpha}\tau$. Since mode frequency fluctuations were the largest temperature-dependent sources of error in Ref.~\cite{srinivas_2021}, AESE could reduce the need for sub-Doppler cooling. In current QCCD algorithms, we spend over an order of magnitude more time on ground state cooling than on two-qubit gates \cite{pino_2021,moses_2023}. Performing gates with AESE could, therefore, lead to reductions in algorithm time even if it increases gate time. \\

\noindent To exemplify this, we examine the effect of a static shift $\varepsilon_{\alpha}$, which gives:
\begin{eqnarray}\label{eq:ham_mot_shift}
\hat{H}_{g}=\hbar\gamma(t)\sum_{\alpha=s,c}\hat{S}_{z,\alpha}\Omega_{\alpha}^{\prime}[\hat{a}_{\alpha}^{\dagger}e^{i\delta_{\alpha}t} +\hat{a}_{\alpha}e^{-i\delta_{\alpha}t}] + \hbar\varepsilon_{\alpha}\hat{a}_{\alpha}^{\dagger}\hat{a}_{\alpha}.
\end{eqnarray}
Following the derivation given in Ref.~\cite{sutherland_2022_1}, again, we transform this equation into a frame that rotates with $\hbar\varepsilon_{\alpha}\hat{a}_{\alpha}^{\dagger}\hat{a}_{\alpha}$:
\begin{eqnarray}\label{eq:ham_mot_shift}
\hat{H}_{g}=\hbar\gamma(t)\sum_{\alpha=s,c}\hat{S}_{z,\alpha}\Omega_{\alpha}^{\prime}[\hat{a}_{\alpha}^{\dagger}e^{i(\delta_{\alpha}+\varepsilon_{\alpha})t} +\hat{a}_{\alpha}e^{-i(\delta_{\alpha}+\varepsilon_{\alpha})t}]. \nonumber
\end{eqnarray}
Without AESE, when $t_{g}$ is not an integer multiple of $2\pi /[\delta_{\alpha}+\varepsilon_{\alpha}]$, the $1^{\mathrm{st}}$ order terms in Eq.~(\ref{eq:magnus}) no longer vanish, and contribute an amount:
\begin{eqnarray}\label{eq:spin_motion_entanglement_error}
    \mathcal{I}_{1} &=& \frac{\pi^{2}}{64}\sum_{\alpha}\frac{\varepsilon^{2}_{\alpha}}{\Omega_{g,\alpha}^{2}}(2n_{\alpha}+1)\lambda_{\hat{S}_{z,\alpha}}^{2} \nonumber \\
    &=& \frac{\pi^{2}}{40}\sum_{\alpha}\frac{\varepsilon^{2}_{\alpha}}{\Omega_{g,\alpha}^{2}}(2n_{\alpha}+1)
\end{eqnarray}
to the infidelity, where we have assumed $\gamma(t)=1$, and, in the second line, averaged over SU(4) to get $\lambda_{\hat{S}_{z,\alpha}}^{2}=8/5$. With AESE, $\mathcal{I}_{1}$ vanishes when the limit $1/\tau\gg \delta_{\alpha}$ is satisfied, independent of the particular value of $\delta_{\alpha}$; as long as this condition remains true, we can assume the contribution from residual spin-motion entanglement to be negligible. \\

\noindent With or without AESE, drifts of the motional frequencies of the gating modes cause the qubits to accumulate entanglement at the wrong rate. This causes a temperature independent error. Assuming the errors from each mode $\alpha$ are additive, we can insert $\omega_{\alpha}\rightarrow \omega_{\alpha}+\varepsilon_{\alpha}$ and $\delta_{\alpha}\rightarrow \delta_{\alpha}+\varepsilon_{\alpha}$, into the time propagator for the gate:
\begin{eqnarray}
    \hat{U}_{t}&=& \prod_{\alpha}\exp\Big(i\braket{\gamma^{2}}t_{g}\hat{S}_{z,\alpha}^{2}\frac{\Omega_{\mathrm{g}}^{\prime 2}\omega_{\alpha}}{[\delta_{\alpha}+\varepsilon_{\alpha}][\omega_{\alpha}+\varepsilon_{\alpha}]} \Big).
\end{eqnarray}
Keeping only leading-order terms in $\varepsilon_{\alpha}/\delta_{\alpha}$ and $\varepsilon_{\alpha}/\omega_{\alpha}$, we can factor $\hat{U}_{t}=\hat{U}_{g}\hat{U}_{\mathrm{e}}$, where $\hat{U}_{g}$ the `ideal' gate operator, and:
\begin{eqnarray}
    \hat{U}_{\mathrm{e}}&=& \prod_{\alpha}\exp\Big(-i\braket{\gamma^{2}}t_{g}\hat{S}_{z,\alpha}^{2}\frac{\varepsilon_{\alpha}\Omega_{g,\alpha}^{2}[\omega_{\alpha}+\delta_{\alpha}]}{\omega_{\alpha}\delta_{\alpha}^{2}}\Big),
\end{eqnarray}
is the error Hamiltonian. Again following the prescription outlined in Ref.~\cite{sutherland_2022_1}, we can immediately compute the expected infidelity:
\begin{eqnarray}\label{eq:infidelity_static}
\mathcal{I}_{\mathrm{m}}&=& \sum_{\alpha}\frac{16\braket{\gamma^{2}}^{2}\Omega_{g,\alpha}^{4}[\omega_{\alpha}+\delta_{\alpha}]^{2}}{5\omega_{\alpha}^{2}\delta_{\alpha}^{4}}\varepsilon_{\alpha}^{2}t_{g}^{2},
\end{eqnarray}
where, in the second  line, we have inserted $\lambda_{\hat{S}_{z}^{2}}^{2}=16/5$ via averaging over SU(4) for $\ket{\psi(0)}$. Importantly, the infidelity due to this error mechanism is independent of the motional state of the ions, which relaxes the requirement for ground state cooling. Plugging in the the parameters for the RF-gradient two-qubit gate exemplified in Sec.~\ref{sec:heating} to Eq.~(\ref{eq:infidelity_static}), we get $\mathcal{I}_{\mathrm{m}}\simeq 10^{-7}$ if we assume $\varepsilon_{\alpha}/2\pi= 50~\mathrm{Hz}$; this is over two orders-of-magnitude smaller than the value of $3\times 10^{-5}$ that was estimated for a ground state cooled crystal experiencing that shift in Ref.~\cite{srinivas_2021}. \\

\noindent In addition to drifts of $\omega_{\alpha}$, there is also a static shift from the Kerr-like interaction between the radial and axial STR modes \cite{roos_2008_kerr, nie_2009}. The size of this shift is proportional to the phonon occupation of that spectator mode, and is typically larger when $\omega_{s}$ is smaller; this means the infidelity associated with it is a potential concern, especially when operating at Doppler temperatures. Fortunately, AESE renders the crystal significantly less sensitive to static motional shifts. Plugging in parameters for the example two-qubit gate discussed in Sec.~\ref{sec:heating}, the formula derived in Ref.~\cite{nie_2009} predicts this shift will be $\varepsilon_{s}/2\pi \simeq -31.5\times  n_{a}~\mathrm{Hz}$, where $n_{a}$ is the number of phonons in the axial STR mode. To analyze this effect at Doppler temperatures, we set $n_{a}=10$, making $\varepsilon_{\alpha}/2\pi \equiv -315~\mathrm{Hz}$, which gives $\mathcal{I}_{\mathrm{m}}\simeq 5\times 10^{-6}$. Finally, we note that the infidelity due to static motional shifts can be suppressed arbitrarily by increasing the values of $\delta_{\alpha}$, giving experimentalists the ability to exchange the size of this error for an increase in $t_{g}$. \\

\subsection{Potential obstacles}\label{sec:issues}

In this section, we anticipate three potential sources of noise that laser-free gates performed with AESE would not necessarily be insensitive to: memory errors, the non-repeatability of gradient field\textemdash both of which can be dynamically decoupled\textemdash and motional frequency fluctuations near-resonant to $\delta_{\alpha}$.

\subsection*{Memory Errors}\label{sec:memory_errors}

The scheme we are describing is a direct $\hat{\sigma}_{z,1}\otimes \hat{\sigma}_{z,2}$ geometric phase gate \cite{leibfried_2003}, which cannot be performed on a magnetic field insensitive `clock' state, meaning the gate is sensitive to memory errors during operation. We take the Hamiltonian representing this error mechanism as:
\begin{eqnarray}\label{eq:memory_ham}
    \hat{H}_{\mathrm{mem}} = \frac{\hbar}{2}\sum_{j}\Delta_{j}(t)\hat{\sigma}_{z,j},
\end{eqnarray}
where $\Delta_{j}\equiv B_{j}(t)\partial \omega_{j}/\partial B_{j}$ is the energy shift of each ion due to an unaccounted for magnetic field acting on the qubit. Crucially, Eq.~(\ref{eq:memory_ham}) commutes with Eq.~(\ref{eq:first_geo}) at all times, meaning we can dynamically decouple the effects of slow $\dot{\Delta}_{j}(t) < 2\pi/t_{g}$ with a spin-echo, or higher-order Walsh sequence \cite{hayes_2012}. Reference~\cite{srinivas_2021} was able to suppress the effect of higher-frequency magnetic field noise on their gate by over an order of magnitude using `Intrinsic Dynamical Decoupling' (IDD) \cite{sutherland_2019,sutherland_2020} via tuning the bichromatic carrier interaction already present in the gate. It is possible to perform gates with AESE and IDD using the $\propto~ J_{2}$ scheme described in Ref.~\cite{sutherland_2019}, but it would require additional control fields, add sources of infidelity due to higher-order interactions with the carrier \cite{roos_2008}, and slow the gate due to the $\propto~ J_{2}$ dependence of the effective spin-dependent force.

\subsection*{Control field fluctuations}\label{sec:control}

In contrast to previous laser-free gating schemes that drive the qubit transition with the gradient \cite{ospelkaus_2008,ospelkaus_2011, harty_2016, zarantonello_2019} or an added microwave field \cite{mintert_2001,weidt_2016,webb_2018,sutherland_2019,srinivas_2021}, Eq.~(\ref{eq:first_geo}) only contains $\propto~ \hat{S}_{z,\alpha}$ terms. This means that the scheme does not necessitate nulling the magnetic field at the point of the ions when it is performed with a spin-echo. This could simplify experiments and allow for larger gradients. For the schemes described above that generate spin-motion coupling with a DC gradient, the gate will be sensitive to the non-repeatability of the gradient during the two halves of the spin-echo. Fortunately, there are several ways of avoiding this problem. First, if we use the RF gradient technique described in Sec.~\ref{sec:rf}, non-repeatability is a non-issue because the gradient will effectively dynamically decouple the effect on a timescale set by $\omega_{g}$. For a DC gradient, this sensitivity could be coherently suppressed with Walsh sequences \cite{hayes_2012}, or by nulling the magnetic field with highly-correlated electrodes such that their fluctuations cancel \cite{hahn_2019}. 

\subsubsection*{Time-dependent fluctuations of the gate mode frequencies}\label{sec:time_dep_fluct} 
\noindent Section~\ref{sec:motional_decoherence} showed that performing gates with AESE significantly reduces sensitivity to static shifts of the gate mode frequencies $\omega_{\alpha}\rightarrow \omega_{\alpha}+\varepsilon_{\alpha}$. This decrease in sensitivity is general so long as the limit $\delta_{\alpha}/2\pi \gg 1/\tau$ remains satisfied. For a time-dependent shift $\omega_{\alpha}\rightarrow \omega_{\alpha}+\varepsilon_{\alpha}(t)$, if the noise-spectral density (NSD) of $\varepsilon_{\alpha}(t)$ has a non-negligible component near $\delta_{\alpha}$ there will be spin-motion coupling terms that do not vanish with AESE, causing residual spin-motion coupling and errors that are sensitive to temperature. \\

\noindent To illustrate this potential issue qualitatively, we examine the infidelity from `motional dephasing' described in Ref.~\cite{sutherland_2022_1}. This derivation assumes that the motional frequency fluctuations $\varepsilon_{\alpha}(t)$ follow a white noise NSD, an assumption that nullifies the coherent suppression of spin-motion entanglement from AESE. Importantly, this assumption is typically chosen to simplify calculations, not because it is an accurate description of the NSD. The NSD for $\varepsilon_{\alpha}(t)$ is usually better described as $\propto~ 1/f_{\alpha}$ noise, meaning lower-frequency drifts\textemdash that enable AESE to suppress spin-motion entanglement\textemdash will be much more likely than spectral components near $\delta_{\alpha}$. Therefore, in order to represent a (plausible but unlikely) worst-case-scenario, we assume white noise and apply the derivation exactly as described in Ref.~\cite{sutherland_2022_1}. This gives:
\begin{eqnarray}
    \mathcal{I}_{\mathrm{d}} &\simeq& \sum_{\alpha}\frac{2\eta_{\alpha}\Omega_{g,\alpha}^{2}t_{g}}{\delta_{\alpha}^{2}}\Big([2n_{\alpha}+1]\lambda_{\hat{S}_{z,\alpha}}^{2} + \frac{3\Omega_{g,\alpha}^{2}}{\delta_{\alpha}^{2}}\lambda_{\hat{S}_{z,\alpha}^{2}}^{2} \Big) \nonumber \\
    &= & \sum_{\alpha}\frac{16\eta_{\alpha}\Omega_{g,\alpha}^{2}t_{g}}{5\delta_{\alpha}^{2}}\Big([2n_{\alpha}+1] + \frac{6\Omega_{g,\alpha}^{2}}{\delta_{\alpha}^{2}} \Big)
\end{eqnarray}
where $\eta_{\alpha}=2/t_{\mathrm{d},\alpha}$, and $t_{\mathrm{d},\alpha}$ is the coherence time between neighboring Fock states of mode $\alpha$. We have, again, averaged over SU(4) in the second line. For the system we introduced in Sec.~\ref{sec:heating}, assuming $n_{s}=n_{c}=10$ to represent Doppler temperatures, we get an infidelity of $\mathcal{I_{\mathrm{d}}}\simeq 3\times 10^{-3}$. This indicates that, although unlikely, if there is a non-trivial component of the motional shift NSD, then it may become necessary to cool the system below the Doppler limit.

\section{Outlook}

The gating protocols analyzed here, while similar to other state-of-the-art approaches, work in a parameter regime that may significantly reduce a number of leading technical challenges when incorporating high-fidelity geometric phase gates into large-scale quantum computer architectures.  In particular, if we drive the gate with an off-resonant spin-dependent force, and suppress any residual entanglement with pulse shaping, we can relax the the gate's temperature requirements.  Our analysis implies we could achieve two-qubit gate errors at or below the $10^{-4}$ level while operating at Doppler temperatures. Eliminating the need for ground state cooling could allow for faster, hotter, transport operations and remove the time overhead associated with ground-state cooling schemes\textemdash which would improve circuit times in QCCD architectures. Finally, the ability to modulate spin-motion couplings via a Rabi or adiabatic rapid passage transitions between clock and Zeeman states is a promising strategy for operating laser-free gates using ferromagnetic gradients.  Performing entangling gates that require neither lasers \emph{nor} the driving of large currents is an enticing prospect for avoiding many of the key technical challenges to scaling trapped ion quantum computers.

\section*{Acknowledgements}
We would like to thank D. H. Slichter, D. Leibfried, J. P. Gaebler, B. J. Bjork, P. J. Lee, C. Langer, E. Hudson, and D. Hayes for useful conversations.

\section*{References}
\bibliography{biblio}
\bibliographystyle{iopart-num}
\end{document}